\documentstyle[prd,aps]{revtex}
\begin{document}
 \input epsf
\draft
\renewcommand{\topfraction}{0.8}
\twocolumn[\hsize\textwidth\columnwidth\hsize\csname
@twocolumnfalse\endcsname
\preprint{CITA-98-33, hep-ph/9808xxx, August, 1998}
\title {\bf   Sine-Gordon Parametric Resonance}
\author{Patrick B. Greene}
\address{Department of Physics, University of Toronto,
60 St George Str, Toronto, ON M5S 1A7, Canada\footnotemark[1]}
\author{Lev Kofman}
\address{ CITA, University of Toronto, 60 St George Str,
Toronto, ON M5S 1A1, Canada\footnotemark[2]}
\author {Alexei A. Starobinsky}
\address{ Landau Institute for Theoretical Physics,
Kosygina St. 2, Moscow 117334, Russia}
\date { \today	}
\maketitle
\begin{abstract}
  We consider the instability of fluctuations  in an  oscillating
scalar  field which obeys  the Sine-Gordon equation.
We present simple closed-form
analytic solutions describing the parametric resonance in the Sine-Gordon
model. The structure of the resonance   differs from that obtained with
the Mathieu equation which is usually derived with the small angle 
approximation to the equation for fluctuations. 
The results are applied to axion cosmology, where the
oscillations of the classical axion field, with a Sine-Gordon
self-interaction potential, constitute the cold dark matter of the  
universe.
When the axion misalignment angle at the QCD epoch, $\theta_0$, 
is small, the parametric
resonance of the axion fluctuations is not significant. However, in
regions of larger $\theta_0$ where axion  miniclusters would form,
the resonance may be important. As a result, axion miniclusters
may disintegrate into finer, denser clumps.
We also apply the theory of Sine-Gordon parametric resonance 
to reheating in the Natural Inflation scenario.  The decay of the inflaton
field due to the self-interaction alone is ineffective,
but a coupling to other bosons  can lead to preheating 
in the broad resonance regime.  Together with the preheating of fermions, 
this can alter the reheating scenario for Natural Inflation.

\end{abstract}
\pacs{PACS: 98.80.Cq  \hskip 2.5 cm CITA-98-33 }
 \vskip2pc]

\section{Introduction}

\footnotetext[1]{On leave from the 
Department of Physics and Astronomy, University of Hawaii,
2505 Correa Rd., Honolulu, HI 96822 USA}
\footnotetext[2]{On leave from the 
Institute for Astronomy, University of Hawaii,
2680 Woodlawn Dr., Honolulu, HI 96822, USA}
An oscillating pseudo-Nambu-Goldstone (pNG) scalar field, $\phi$,
governed by the Sine-Gordon theory
and the parametric resonance of its fluctuations,
$\phi_k e^{i \vec k \vec x}$,
is a prototype for several interesting cosmological and other applications.
One of these is the cosmic axion.
The axion field, which was originally invented in order to solve the strong CP
problem \cite{pq}, is known to be a dark matter
candidate \cite{axion1,axion2,axion3},
for  reviews  see e.g. \cite{ax,KT,strax}.
In the literature, it is often assumed that the 
axion field is a relatively homogeneous, oscillating classical scalar  
field which  does not decay to other particles. 
The possibility of a parametric resonance decay of this field due to its 
self-interaction was considered in Ref. \cite{axion2}.
Using the small angle approximation, which leads to a
Mathieu equation, and  an elementary estimate  for the width
of the instability band and the redshifting of the modes
due to the expansion of the universe,
it was concluded that this process is inefficient.
Another important application of the Sine-Gordon theory is to Natural  
Inflation \cite{natural}, where after inflation a pseudo-Goldstone field
oscillates coherently about the minimum of its
Sine-Gordon potential.  The parametric resonance of inflaton
fluctuations due to the self-interaction in the
natural inflation scenario was considered in \cite{df}. Again, using the
small angle approximation and the  Mathieu equation, it was concluded that
the resonance  is not effective. One more  application is the
resonant decay of disordered   chiral condensates \cite{dcc},
modeled with the Sine-Gordon equation \cite{belova}.

In the last few years, it has been appreciated that parametric  
resonance of coherently oscillating scalar fields, in particular, 
the inflaton field (preheating) can be very 
efficient \cite{KLS}.  For this reason, axion  
parametric resonance was recently re-examined  \cite{KSS}.
The   equations governing axion fluctuations were solved
 numerically in Minkowski space-time and in an expanding universe
 under  the simplifying assumption  that the axion mass is a non-vanishing
constant since the very beginning of the radiation dominated universe.  
The results confirmed  the conventional wisdom
that axion parametric resonance is inefficient  because of the expansion
 of the universe. However, in the cosmic axion scenario,
 the axion mass is initially absent;  it turns on very abruptly at 
the QCD epoch, $t_{QCD}$, when instanton effects break the classical
Peccei-Quinn symmetry, $U_{PQ}(1)$.  After $t_{QCD}$,
the axion begins to oscillate with a period many orders of
magnitude smaller than the age of the universe.  
This is different from the setting of ref.~\cite{KSS}, 
where consequently, the effect of expansion is overestimated.

In this paper we will systematically study the parametric resonance
in the Sine-Gordon theory, both in Minkowski space-time and
in an expanding universe. We will show how this problem can be 
solved analytically and present
the theory of the  parametric resonance of the Sine-Gordon equation.
For this, we will use  methods which we elaborated
in application to the theory of preheating after inflation \cite{KLS97,GKLS}.
With these novel tools, we will reexamine the axion parametric resonance
in an expanding universe where the axion mass is switched on at $t_{QCD}$.
It turns out that the results of the Sine-Gordon
parametric resonance  can be extended to axions in an expanding universe
analytically.
In agreement with \cite{axion2,KSS} we will
show that the  axion parametric resonance is ineffective
as soon as the axion misalignment angle at the QCD  epoch, $\theta_0$, 
is small,  $\theta_{0} {\sim} 10^{-3}$.

However, there are axion scenarios where the axion field $\theta_0$
is  spatially inhomogeneous, and in  its peak regions
can be significantly larger than its mean amplitude.
When cosmic axions become gravitationally dominant, these regions
will turn into  dense axion miniclusters \cite{miniclusters,ktk}.
We consider the axion resonance in regions  where the coherently 
oscillating  axion field of the large amplitude $\theta_0$ is
 smooth at the scales of  proto-miniclusters. In these regions,
 the axion parametric resonance can be quite significant
and leads to the decay of the smooth axion field into
an inhomogeneous field within  proto-miniclusters.
Then, when axions become gravitationally dominant, these
inhomogeneities  will form small high density  clumps
inside the minicluster.

Finally, we turn to another application of the
Sine-Gordon parametric resonance: reheating in natural inflation.
We consider the parametric resonance of inflaton fluctuations
due to the self-interaction.
We confirm the result of  \cite{df}, although by different reasoning,
that parametric resonance due to the self-interaction
is not effective.  For the sake of completeness, 
we also consider the creation of other bosons $\chi$
via a coupling of the form $g^2\phi^2\chi^2$ with the inflaton.
In this case, $\chi$-particles can be created in the
broad parametric resonance regime; and thus, preheating in natural
inflation can be very efficient.

\section{\label{solution} Parametric Resonance in the Sine-Gordon Theory}

First, we neglect expansion of the universe ($a=1$, $H = {{\dot a}
\over a} = 0$), and consider the parametric resonance in the
 Sine-Gordon theory. We  will start with
 the time evolution of the homogeneous   background oscillations
in this theory.  Then
 we turn to the basic issue of our study, the stability of
small inhomogeneous fluctuations $\phi_k(t)e^{i{\vec k \cdot \vec x}}$
of comoving momentum  ${\vec k}$.
Since the background oscillations are
periodic,  the temporal part of the
eigenfunction is $\phi_k(t) \sim e^{\mu_k mt}$, where
 the characteristic exponent $\mu_k$  is real for resonant modes.
The goal of this Section is to find $\mu_k$ in the Sine-Gordon theory.
Fortunately,
one can avoid making any further simplification (besides
neglecting expansion)
because it is possible to find   exact resonance
solutions   using the methods
 developed in Ref. \cite{GKLS}.
Once we have developed the solutions,
we compare the exact result with the estimation
obtained from the small angle approximation, $\theta_0 \ll 1$, 
which leads to a Mathieu equation for the fluctuations.

\subsection{\label{SGbackground}
 Background Oscillations in the Sine-Gordon Theory}

\begin{figure}[t]
\centering
\leavevmode\epsfysize=5.3cm \epsfbox{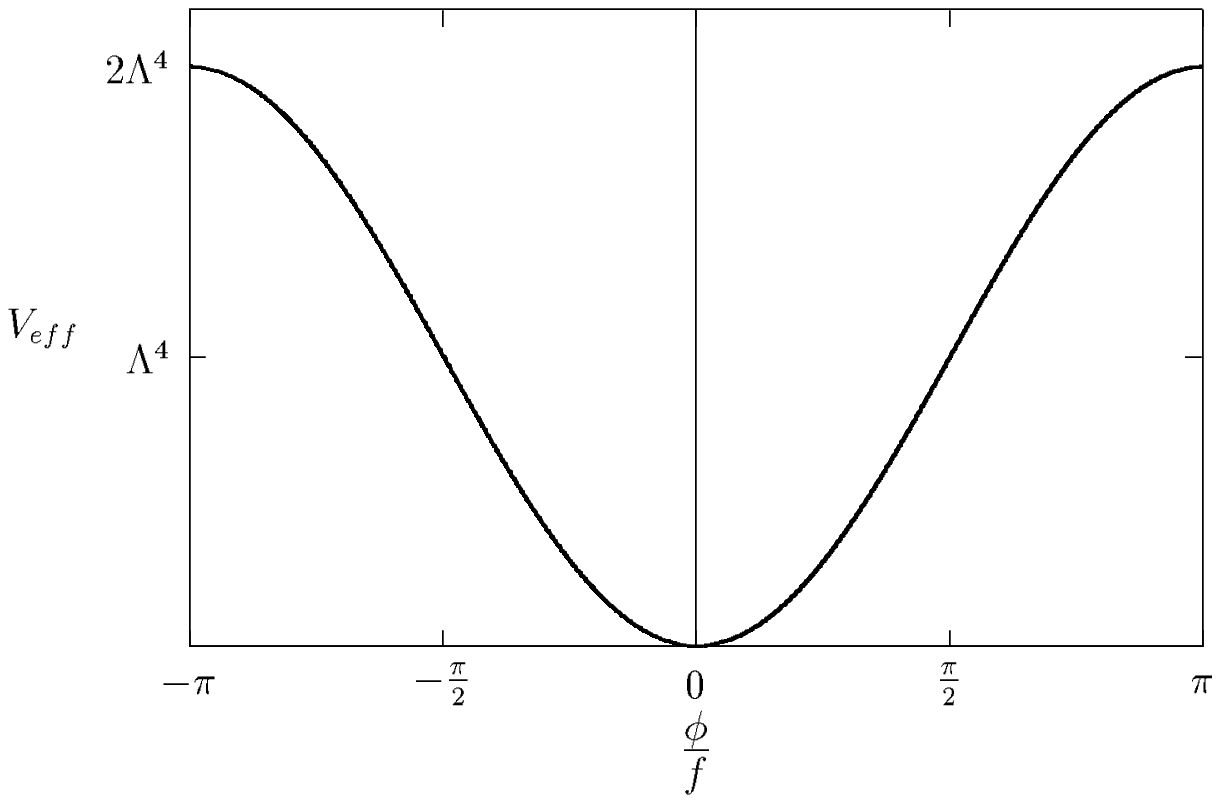}\\

\caption[fig1]{\label{fig:fig1} The potential
$\Lambda^4 {\Bigl(1-\cos{\frac{\phi}{f}}\Bigr)}$.}
\end{figure}

The potential of the scalar field in the Sine-Gordon   theory
  can be represented as follows:
\begin{equation} \label{1}
V(\phi) =  \Lambda^4 \left (1 - \cos {\phi\over f  }\right ) \ ,
\end{equation}
where $\Lambda^4 = m^2 f^2$ with
$f$ a constant (corresponding to the radius of the tilted
``wine bottle'' potential) and
$m$ the mass of the bose field $\phi$,
see Fig.~\ref{fig:fig1}.
To simplify notation,   we will sometimes use the dimensionless
misalignment angle,  $\theta \equiv {\phi \over f}$.

In general,  the   field $\phi(t, {\vec x})$ obeys the
  Sine-Gordon equation. For a homogeneous
scalar field    $\phi = \phi(t)$ and without expansion of the  
universe, we have
\begin{equation}\label{SG}
\ddot \phi + m^2 f \sin {\phi \over f}=0 \ ,
\end{equation}
where $m$ is a constant mass and $f$ is the radius of the potential.
For small initial misalignment angle, $\theta_0 \equiv {\phi_0 \over  
f} \ll 1$,
the solutions to this equation are approximately harmonic.
Nevertheless, we will derive the general  solution  of this equation
for an arbitrary initial amplitude without the
assumption of a small misalignment angle.
Although, for practical purposes in dealing with axion parametric  
resonance
one may skip this general consideration, it is instructive
to consider parametric resonance in the Sine-Gordon theory for a wide
range of parameters. This will eventually  allow us to understand how
the parametric resonance in an {\em expanding} universe can be so  
sensitive
to the initial conditions and parameters involved.

Without loss of generality, we can restrict ourselves to the finite
 motion $\vert \phi(t) \vert \leq
\pi f$. The energy integral of Eq.~(\ref{SG}) is
\begin{equation}\label{energy}
{\dot \phi^2 \over 2} + V(\phi) = E \ ,
\end{equation}
where $E \leq 2 m^2 f^2$, $V(\phi)$ is given by Eq.~(\ref{1}),
and $E= V(\phi_0)$.

In Appendix A we derive the final result for the evolution of the
background oscillations
\begin{equation}\label{hom}
\phi(t)=2f \,
 \arctan \left( \sqrt{\epsilon \over {2 - \epsilon}}
cn ( mt; \, {\cal K}) \right)  \ ,
\end{equation}
where the useful  dimensionless parameter is
$\epsilon \equiv {E \over \Lambda^4} = 1-\cos{\theta_0}$ and
the modulus ${\cal K}=\sqrt{E \over {2 \Lambda^4}}=\sqrt{ \epsilon  
\over 2}$.
As follows from Eq.~(\ref{hom}), it is convenient to use  the
dimensionless time variable
\begin{equation}\label{time}
\tau=  mt \, .
\end{equation}
As shown in Appendix A, the period of the background fluctuations
in the dimensionless time variable,  
$T_{\tau}$, can be expressed through  a complete
elliptic integral of the first kind:
$T_{\tau}=4{\bf K} \Bigl( \sqrt{ \epsilon \over 2 }\Bigr) $.
One may check that the amplitude of $\phi(t)$ from Eq.~(\ref{hom}) is
$\phi_0$.

In the small amplitude limit, $ \phi_0 \ll f$, we have
$E \approx {m^2 \phi_0^2 \over  2}$ and  the oscillations
of the background field are approximately harmonic:
$\phi(t) \approx \phi_0  \cos{mt}$, $T_{\tau} \approx 2\pi$.

\subsection{\label{solution2} Analytic Resonant Solutions}

We shall consider the fluctuations $\phi_k(t)$ in the Sine-Gordon model
without expansion of the universe, but not assuming the misalignment angle
${\phi_0 \over f}$ to be small. In this case the mode equation for
fluctuations is
\begin{equation}\label{sgf}
\ddot \phi_k+{\left(k^2 + m^2\, \cos {\phi \over f}
\right)} \phi_k  = 0 \ .
\end{equation}
Let us  use the  dimensionless time, $\tau=mt$,
and the  dimensionless momentum $\kappa \equiv { k \over  m}$.
Then the mode equation is
\begin{equation}
\ddot \phi_k  +  {\left(\kappa^2  + \cos {\phi(\tau) \over f}
 \right)} \phi_k  = 0 \ ,
\label{fluc3}
\end{equation}
where the time-dependence of the background field, $\phi(\tau)$,
is given by Eq.~(\ref{hom}).  Recall that
$\phi(\tau)$ is a periodic function of $\tau$ with period $T_{\tau}$.

Equation  (\ref{fluc3})  describes an
oscillator, $\phi_k(\tau)$,    with a variable frequency which
depends periodically on time.  As is well known, the solutions  
$\phi_k(\tau)$
can be exponentially unstable,
$\phi_k \propto e^{ \mu_k \tau}$, for some ranges of the
parameters $\kappa$ and $\phi_0 \over f$.
For positive-frequency vacuum initial conditions,
 $ \phi_k(\tau) \simeq { e^{-i\kappa \tau } \over \sqrt{2\kappa} }$,
one expects  the exponentially fast
creation of 
particles 
$n_k \propto  e^{2 \mu_k \tau}$
in the unstable modes as the background  field oscillates.
For a given mode, $\kappa$, the strength of interaction is  
determined by the
dimensionless background amplitude, or misalignment angle,
$\theta_0={\phi_0 \over f}$.
That is, the absolute value of the misalignment angle, $\theta_0$,  
ultimately defines the structure of the parametric 
resonance in Sine-Gordon theory.
It turns out that the strength of the resonance monotonically
depends on this parameter; it is stronger for larger
 $\vert \theta_0 \vert $. For $\vert \theta_0 \vert \geq \pi/2$
the effective mass is negative, and one has even  a strong
 tachyonic instability.

The main purpose of our study in this Section is to find
the characteristic exponent $\mu_k=\mu_k (\theta_0)$.
Even neglecting the expansion of the universe,  the
equation for fluctuations (\ref{fluc3})
 looks difficult   at first glance.
Fortunately, it turns out that in the Sine-Gordon theory
one can obtain simple, closed-form analytic solutions to
the mode equation  (\ref{fluc3}).

For this, let us rewrite this equation in a different form.
We will use a new ``time''  variable $z$:
\begin{equation}
z(\tau) \equiv \cos{\phi(\tau) \over f} \ ,~~\
{d \over d\tau}= -\sqrt{2(1-z^2)(\epsilon-1+z)}
  {d \over dz} \ ,
\label{z}
\end{equation}
where, as before, $\epsilon={E \over \Lambda^4}=1-\cos{\theta_0}$.
Note that $z$ is sandwiched in the range  $1-\epsilon \leq z \leq 1$.
Equation  (\ref{fluc3}) for
fluctuations becomes
\begin{eqnarray}
&&2(1-z^2) (\epsilon-1+z) \phi_k'' +
\Bigl(1+2(1-\epsilon)z-3z^2 \Bigr)\phi_k'
 \nonumber\\
&&+( \kappa^2 + z)\phi_k=0 \ ,
\label{fluc4}
\end{eqnarray}
where $(...)'$ stands for the $z$-derivative.
As we show in Appendix B, the mode equation in this form
coincides with  the algebraic form
of the Lam\'{e} equation with particular coefficients.

At this point we will repeat the trick we found in \cite{GKLS}
for a similar problem in $\lambda \phi^4$ theory.
Specifically, we introduce  two linearly-independent solutions of
Eq.~(\ref{fluc4}), $\phi_{1}(z)$ and $\phi_{2}(z)$,
where the lower index $k$ is omitted for simplicity.
Let us  construct  the bilinear combinations
$\phi_1^2$, $\phi_2^2$, and $\phi_1 \phi_2$. From (\ref{fluc4})
 it follows that
these bilinear combinations obey a third order equation
\begin{eqnarray}
&&2(z^2-1)(\epsilon-1+z)  M''' +
\left[9z^2-6(1-\epsilon)z -3\right] M''  \nonumber \\
&&+ 2(z-1+ \epsilon
     -2 \kappa^2)M' -2 M =0 \ .
\label{polin}
\end{eqnarray}
The three solutions, $M(z)$, of this equation  correspond to
(up to a numerical normalization ${\cal N}$ still to be given)
the three bilinear combinations of $\phi_1$ and $\phi_2$.

Equation (\ref{polin}) admits a polynomial solution
\begin{equation}\label{deg}
M(z)=z-1+ \epsilon-2\kappa^2 \ .
\end{equation}
In the resonance zone,
this polynomial solution must be the product of an
exponentially growing solution and an exponentially decreasing one, i.e.
\begin{equation}\label{n1}
\phi_1(z) \phi_2(z)={\cal N}^2 \cdot ( z-1+ \epsilon-2\kappa^2) \ ,
\end{equation}
where $\cal N$ is a normalization factor.
An additional relation between the two fundamental solutions is given by
the Wronskian of Eq.~(\ref{fluc4})
\begin{equation}\label{W}
\phi_1  \phi_2' -
\phi_2'   \phi_1 ={\cal N}^2 { c_k \over
 \sqrt{(1-z^2)(\epsilon-1+z) }} \ ,
\end{equation}
where $c_k$ is a constant to be determined.
From  (\ref{n1}) and   (\ref{W}) we  obtain the closed
form analytic solutions
\begin{eqnarray}
&&\phi_{1,2}(z) ={\cal N} \sqrt{ \vert M(z)  \vert} \times  \nonumber\\
&& \exp \left( \pm {c_k \over 2} \int { dz \over
 {\sqrt{(1-z^2)(\epsilon-1+z)} M(z) }} \right) \ .
\label{form1}
\end{eqnarray}
The solution to Eq.~(\ref{fluc3}) with positive-frequency vacuum 
initial condition will be a linear superposition of the two fundamental
solutions $\phi_{1,2}$ \cite{kaiser98}.
The amplitude of fluctuations is rapidly dominated by the growing term.
As is shown in Appendix C, we can extract the characteristic exponent
describing the resonance effect from the fundamental solutions
themselves.  Therefore we will not further  discuss the numerical
normalization and, for simplicity, we normalize our fundamental solutions
to unit initial amplitude: ${\cal N}=\vert M(1) \vert^{-1/2}$.

\subsection{\label{strength} The Width and Strength of the
  Resonance}

The closed-form analytic resonant solution (\ref{form1})
allows us to find the structure of the resonance in terms of the width
of the resonance band $\Delta \kappa^2$ and its  strength
$\mu_k$ for an arbitrary choice of parameters.
Indeed, substituting this solution back into equation  (\ref{fluc4})
for   $\phi(z)$, we find the constant $c_k$:
\begin{equation}
c_k^2= 8~\kappa^2 \Bigl({\epsilon \over 2} -\kappa^2\Bigr)
\Bigl(1-{\epsilon \over 2} +\kappa^2 \Bigr) \ .
\label{C}
\end{equation}
For exponentially growing solutions, $c_k$ must be real.  Therefore, the
exponentially growing solutions for fluctuations
with $\kappa^2 > 0$
occur in a single instability band for which
$0 \leq \kappa^2 \leq {\epsilon \over 2}$.  In terms of the
basic parameters, the width of the instability band is
\begin{equation}
0 <  \kappa^2 < { {1-\cos{\theta_0}} \over 2}  \ .
\label{instab}
\end{equation}
The growing  solution of Eq.~(\ref{fluc3}) has the form
$\phi_k(\tau)= e^{\mu_k \tau} P[z(\tau)]$, where $P[z(\tau)]$
is a periodic function of  time $\tau $. Using Eq.~(\ref{form1}),
we can now find the characteristic exponent  $\mu_k $
as a function of $\kappa$ for a given initial misalignment angle,
$\theta_0$.  The technical details can be found
in Appendix C.  The result is
\begin{equation}
 \mu_k(\theta_0) = {I_{\kappa} \over T_{\tau}}  \,
\sqrt{2\kappa^2 \Bigl(1-(\cos \theta_0 + 2 \kappa^2)^2 \Bigr) }  \ ,
\label{mu2}
\end{equation}
where an auxiliary  function  $ I_{\kappa}=I_{\kappa}(\theta_0)$ is
\begin{eqnarray}
&&I_{\kappa}(\theta_0)= 4\int\limits_0^{\pi/2} {{d \vartheta}
 \over {\sqrt{ (1+\sin^2 \vartheta)(1+
\cos \theta_0\sin^2 \vartheta)} }} \times \nonumber\\
&&{{\sin^2 \vartheta} \over
{ 1+(\cos \theta_0+2\kappa^2)\sin^2 \vartheta} } \ .
\label{I}
\end{eqnarray}
Recall that $T_{\tau}(\theta_0)=4{\bf K} \Bigl(\vert \sin{ \theta_0 \over
2}\vert
\Bigr)$ is  the period of the background oscillations.

\begin{figure}[t]
\centering
 \hskip -0.5 cm
\leavevmode\epsfysize= 5.5cm \epsfbox{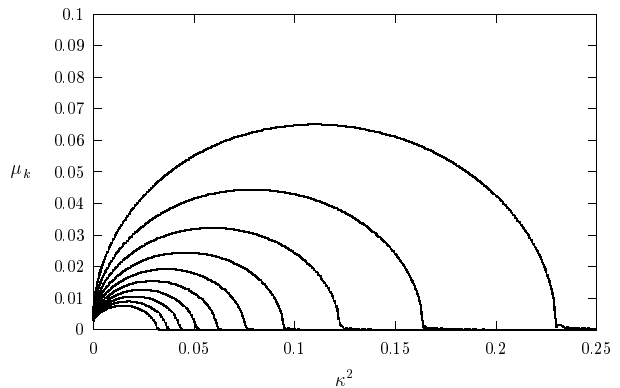}\\
\caption[fig1]{\label{fig:mu}
The characteristic exponent $\mu_k$  of the growth of fluctuations
($\phi_k \propto e^{\mu_k \tau}$) as a function of the (scaled) resonant
momentum $\kappa$ in Sine-Gordon theory. Different curves correspond to
different values of the initial background misalignment angle, $\theta_0$.
{}From the outer curve inwards $\theta_0= 1.67, 1.25, 1.0, 0.83, 0.71,$
$ 0.625, 0.56, 0.5, 0.45, 0.42, 0.38, 0.36.$ }
\end{figure}

In summary, we have found that for an arbitrary value of
$\theta_0$ there
 is always a single resonance band given by formula (\ref{instab}).
The characteristic exponent strongly depends on the amplitude
$\theta_0$ and its value is given by an explicit integral in
formula (\ref{mu2}).
The characteristic exponent $\mu_k$ found by numerical integration of
the equation for fluctuations (\ref{fluc3}) with vacuum positive-frequency
initial conditions is  given in Fig.~\ref{fig:mu} as a function of  
$\kappa$
for various values of the initial background misalignment angle   
$\theta_0$.
These results coincide precisely with the analytic formula (\ref{mu2}).

Let us now consider the case of small misalignment
angle  ${\theta_0} \ll 1$, which is very important for applications.
In the small angle limit, the general formula
 (\ref{mu2}) is reduced to the very simple expression
$\mu_k={\kappa \over 4}\sqrt{ {\theta_0^2   } - { 4\kappa^2}}$;
or explicitly in terms of the momentum $k$,
\begin{equation}\label{simple}
\mu_k={k \over 4m}\sqrt{ {\theta_0^2   } -
  { 4k^2 \over m^2}} \ .
\end{equation}
In this case, the  maximum value $\mu_k ={ \theta_0^2\over 16 }$
occurs at  ${k^2 \over m^2} = { \theta_0^2\over 8 }$.
The resonance band in this limit is
$0  \leq {k^2 \over m^2} \leq { \theta_0^2\over 4 }$.
In the next Section we will extend  formula
(\ref{simple})  taking into account
the expansion of the universe. In this form it
will be used to study axion parametric resonance,
where the background axion misalignment angle will be small
by the QCD time, $t_{QCD}$.

\subsection{\label{solution1} On the Validity of Use of the Mathieu  
Equation}

Let us return to the equation for fluctuations (\ref{fluc3})
and consider a small misalignment angle, $\theta = {\phi \over f} \ll 1$.
There we will  make the approximation:
$\cos{\theta(\tau)} \approx 1 - {1\over 2}\theta^2(\tau) \approx
1-{1\over 2} \theta_o^2 \cos^2{\tau}$.
Thus, to this order in $\theta^2$, Eq.~(\ref{fluc3}) becomes
\begin{equation}\label{math}
\ddot \phi_k    + \left[ \kappa^2
 + \left(1- {\theta_0^2\over 4}\right) - {\theta_0^2\over 4} \cos{(2\tau)}
\right] \phi_k=0 \ ,
\end{equation}
which is a form of the Mathieu equation,
$\ddot \phi_k + \left( A -2q \cos 2 \tau \right)\phi_k=0$, 
 with parameters $A = 1- 2q +
\kappa^2$ and $q = {\theta_0^2\over 8} \ll 1$.  The properties of  
the Mathieu
equation for small $q$ parameter are well known \cite{Mac}.  The resonance
in the first (leading) band  occurs for $A = 1 \pm
q$, which gives a width for the resonance band
${\theta_0^2\over 8} \leq {\kappa^2} \leq
 {3\theta_0^2\over 8} $.
The characteristic exponent in  the first zone is
$\mu_k^{M}=\sqrt{({q \over 2})^2 - (\sqrt{A}-1)^2}$ 
or, explicitly to this order,
\begin{equation}\label{math1}
\mu_k^{M}= \sqrt{ \Bigl( {\theta_0^2 \over 16} \Bigr)^2 -
 \Bigl( {k^2 \over {2m^2}} - {\theta_0^2 \over 8 }  \Bigr)^2   } \ .
\end{equation}
Note that this expression differs from  estimations in  \cite{axion2}.
In the Mathieu approximation the leading resonance band is
concentrated  at ${k^2 \simeq {\theta_0^2\over 4} m^2}$.

We presented here this analysis because it is simple and intuitive.
The approximate solution obtained  with the Mathieu equation
has some qualitative  features similar to the exact solution, 
but it is quantitatively  different. 
In particular, the actual resonance band begins at $k = 0$. 
Even though this band may be 
very narrow, the redshifted resonant modes $\phi_k$ never shift below 
this band,
i.e. expansion of the universe is not as destructive as one might expect.
In this respect, the actual resonance solution is {\it qualitatively}
different from the approximation based on the Mathieu equation.

\section{\label{background}   Background Axion  Field  }

In this and the next  section we will investigate axion
parametric resonance using the results derived above.
Before considering axion fluctuations, we shall discuss the  
evolution of the
background axion field.
In the context of the problem of fluctuations,
it can be either homogeneous or
quasi-homogeneous depending on the cosmic axion scenario
and relation of the length scales of fluctuations to
the scale where the background field is smooth.

The axion, $\phi$, is a pseudo-Nambu-Goldstone boson which appears
after spontaneous symmetry breaking in a theory of a complex scalar field.
It acquires a small mass, $m_a$, due to instanton effects
when the temperature of the universe drops below the QCD scale
\cite{axion1,axion2,axion3}. The effective potential in the simplest
theory of the axion field  can be represented as (\ref{1}),
where $f$ is  the radius of the axion potential
\cite{pq}.
The axion is massless at very high temperature but gradually  
develops a mass.
For $T \gg  \Lambda$ we have $m(T) \approx
0.1 m_a (\Lambda/T)^{3.7}$.
Around the QCD phase transition
($T  \approx \Lambda \approx 200$~MeV), the
mass is  quickly saturated at its low-temperature value,
 $m_a=(\Lambda^2/ f)$, or
 $m_a \approx 0.6
\times10^{-5}\left({10^{12}{\rm Gev}\over f}\right)$ eV.

At this point, the further discussion of the background axion field
branches into different  axion scenarios:
the initial axion misalignment is globally  homogeneous,
the initial  misalignment is homogeneous within a
Hubble domain at the `defrost' moment $t_1$, or
the axions are produced by the decay of axionic strings and walls.
To investigate the axion parametric resonance due to the self-interaction
of the fluctuations with the background axion oscillations, in principle,
one needs to specify the background axion field $\phi(t, {\vec x})$
in these models. However, as we will see below, the axion parametric
resonance depends on the value of $\theta_0$ at the moment of the QCD phase  
transition. As soon as a region with given
misalignment angle $\theta_0$ is smooth
and much larger in size than the typical resonant wavelength,
one can split the axion field $\phi(t, {\vec x})$
into a background and fluctuations around it.

\begin{figure}[t]
\centering
 \hskip -0.5 cm
\leavevmode\epsfysize= 5.5cm \epsfbox{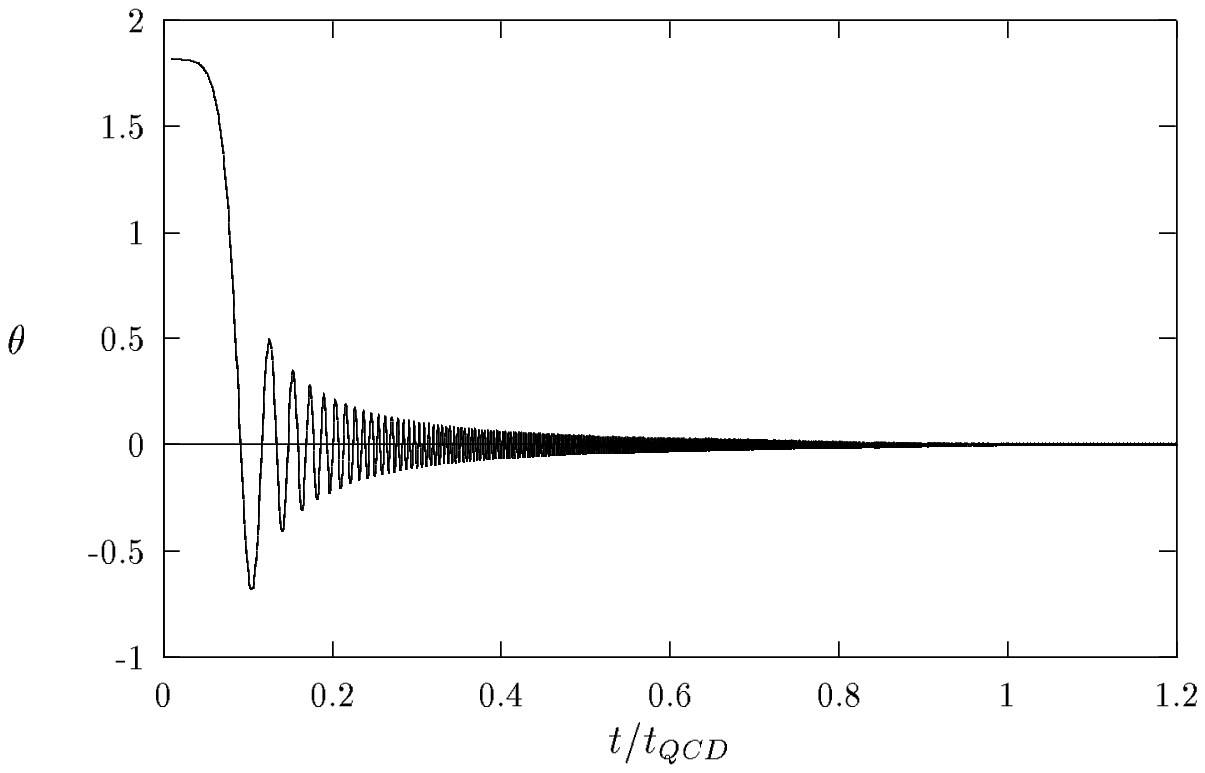}\\
\caption[fig1]{\label{fig:osc}
 Time evolution of the background field $\phi(t)$,
 in terms of the misalignment angle $\theta(t)= {\phi(t) \over f}$,
obtained with the numerical solution of Eq.~(\ref{2}).
The initial value $\theta_1$ is chosen, for illustration, as the
r.m.s. of the random phase, i.e. $\theta_0 = \sqrt{ \pi^2 \over 3} $. }
\end{figure}

We first consider the simple model where $\phi(t)$ is a coherent
homogeneous field throughout the universe, and then will discuss
the inhomogeneous background field $\phi(t,\vec x)$. A homogeneous
axion field obeys the Sine-Gordon equation in an expanding universe
\begin{equation}\label{2}
\ddot \phi + 3H \dot \phi +
 m^2(T) f  \sin   {\phi \over f} =0 \ .
\end{equation}
The time evolution of the background axion field is
plotted in Fig.~\ref{fig:osc}. Before the condition $m(T) \approx 3H(t)$
holds at a moment $t_1$, the axion field is frozen at its initial
 value $\phi_1$ (corresponding to a misalignment angle $\theta_1$).
Shortly after this `defrost' moment $t_1$, the field begins to
oscillate with gradually decreasing amplitude.
 After the axion mass is saturated at  $t_{QCD}$,
the coherent fluctuations are approximately harmonic,
$\phi(t) \approx \phi_0 { \sin m_a t \over a^{3/2}(t)}$,
where $\phi_{0}$ is the amplitude at $t_{QCD}$,
and   $a=\sqrt{t/t_{QCD}}$.

The resonance regime for axion fluctuations starts at $t_{QCD}$
(the growth of fluctuations is negligible prior to that moment).
Therefore, we have to evaluate the misalignment angle
$\theta_{0}$ at   $t_{QCD}$.
If the mass density of coherently oscillating axions
in units of the critical density is $\Omega_a$, then
\begin{equation}\label{omega1}
\theta_0^2 \simeq 10^{-6} \Omega_a h^2 \ ,
\end{equation}
where $h$ is the present Hubble parameter in units of
$100\hbox{\ km/s}\cdot \hbox{Mpc}$.
We assume a temperature at $t_{QCD}$: $T_{QCD} \simeq \Lambda_{QCD} \simeq
{200\hbox{\ MeV}}$.
The value given by Eq.~(\ref{omega1}) is very small and this,
we will see, suppresses the axion resonance.

There are, however, cosmic axion scenarios in which the axion field at the
QCD epoch is globally inhomogeneous.
In this case, the free axion field $\phi(t, {\vec x})$ obeys the
Sine-Gordon equation (\ref{2}) with an additional gradient term
${1 \over a^2} \nabla^2_{\vec x}\phi$.
In   the uncorrelated misalignment scenario \cite{KT,miniclusters}
 there are rare (overdensity) domains where the initial
 phase $\theta_1$ is as large as
$\pi$.
 As was shown in \cite{miniclusters},
large inhomogeneities of the background axion field at the  
present-day scales
$t_1\times (1+z_1) \simeq 10^{18} $cm upwards  result in large
density fluctuations by the epoch when axions gravitationally dominate.
This leads to the formation of axion miniclusters
of the mass $\sim 10^{-12}M_\odot$. For our purpose, it will be 
important that the misalignment angle at $t_{QCD}$ in the 
overdensity domains can be significantly larger than
the average given by  (\ref{omega1}).
The size, $L$, of these domains, or {\em proto-miniclusters},
at $t_{QCD}$ is given by the
size of the horizon at the defrost moment, $t_1$, stretched by the
scale factor $a(t_{QCD})/a(t_1)$. This size $L$
 is smaller than the size of the
horizon at $t_{QCD}$, but significantly larger that
the wavelengths $\sim (\theta_0 m_a)^{-1}$ of axion fluctuations  
subject to
resonant amplification. Therefore, one can consider the resonance of axion
fluctuations around a (locally) homogeneous and coherently  
oscillating field
$\theta$ with an amplitude considerably greater than the estimate
(\ref{omega1}).

The axion field is also inhomogeneous
in the scenario where axions are radiated by axionic cosmic strings
\cite{davis,strax}. However, the history
 of $\theta(t)$ depicted in Fig.~\ref{fig:osc}
is not applicable to this scenario.
Again, the mean value of $\theta_0$ is given by formula (\ref{omega1}),
but the axion field at $t_{QCD}$ is inhomogeneous.
Let us briefly discuss the properties of  $\theta_0 ({\vec x})$
in this scenario.

In this scenario, according to Refs.~\cite{strax},
 the axion field appears after
the universe cools down and PQ symmetry breaking occurs.
In different parts of the universe, on the  order of the size of the
horizon at that  moment, the axion field  takes 
different values in the range from $0$ to $2\pi f$.
As a result, an axionic string network forms. Starting with the moment
 $t_* \sim 10^{-12}\left({m_a \over 10^{-4}\hbox{\ eV}}\right)^4$sec,
the strings begin to oscillate and radiate
their energy into relativistic axions.
The dominant contribution comes from the
smallest loops at a given time which is
some fraction $\alpha$ of the horizon $t$,  $\alpha \sim \kappa$,
where $\kappa \simeq 0.15$ is the backreaction scale (in units of $t$).
The axion spectrum is dominated by the shortest wavelength $\sim  
\alpha t$.  After the axion mass
switches on  at $t_1$, strings transform into domain walls
at the moment $t_w \approx 1.55~ t_1$, when further axion emission
is terminated. Free axions became non-relativistic, with a conserved
occupation number in each mode.
Beginning at the moment  $t_w$, the net  free axion field $\phi(t,  
{\vec x})$ obeys the Sine-Gordon equation. As initial conditions
at $t=t_w$, we take  $\phi(t, {\vec x})$ to be a superposition of
axion waves with uncorrelated phases, emitted by a number of loops,
with the spectrum peaked at the mode $k_w \sim (\alpha t_w)^{-1}$.
In the first approximation, $\phi(t, {\vec x})$ can be modeled by a
random Gaussian field with this spectrum.
Since we will be interested in the parametric resonance
at $t \geq t_{QCD}$ of the modes
$k$ which are many orders of magnitude larger than $k_w$,
we will model the background field
  $\phi(t, {\vec x})$ as a quasi-homogeneous field, with the
scale of the inhomogeneity $L \sim \alpha t_w a(t_{QCD})/a(t_w)$.
The peaks of  the random field  $\theta_0(\vec x)$
can have a field value significantly exceeding the estimation (\ref{omega1}).
Since $k_w \ll m_a$,  for the epoch $t \geq t_{QCD}$ we can neglect
the gradient term in the Sine-Gordon equation for the background field.
At the scales $L$ over which the net axion field  $\phi(t, {\vec x})$
can be considered to be homogeneous, it oscillates with the
frequency $m_a$.  This is an approximation to the actual  
oscillations, which may be neither perfectly coherent nor perfectly harmonic.

\section{\label{resonance}  Axion Parametric Resonance}

\subsection{\label{equations} Axion Fluctuations}

We represent small  fluctuations of the axion field by the modes
$\phi_k(t)e^{i{\vec k \cdot \vec x}}$,  where
${\vec k}$ is the comoving momentum.  These fluctuations, $\phi_k(t)$,
obey  the equation
\begin{equation}\label{axfluc}
\ddot \phi_k + 3H\dot \phi_k + \left({k^2 \over a^2}
 +m_a^2 \cos \theta \right) \phi_k=0 \ .
\end{equation}
One might expect the parametric instability of the fluctuations
due to the quasi-periodic variation of the effective frequency in
Eq.~(\ref{axfluc}).  However, one should also wonder whether  the
expansion of the universe might suppress this effect \cite{axion2,KSS}.

The period of axion oscillations \\
${2\pi \over m_a}  \approx
8 \times 10^{-10} \left({ 5 \cdot10^{-4} \hbox{\ eV} \over m_a} \right)$s; 
meanwhile, the typical time for expansion of the Universe
is $H^{-1} \simeq t_{QCD}
\approx 7.7 \times 10^{-6}$s.  Therefore,
in the first approximation one can neglect the expansion
 of the universe in Eq.~(\ref{axfluc}),  and calculate
 the characteristic exponent $\mu_k$ for the growth of
fluctuations, $\phi_k \propto e^{\mu_k m_a t}$. Then, one
can take into account the  adiabatic  time dependence  of $\mu_k(t)$ and
calculate the net effect as $\phi_k \propto \exp \int_{t_{QCD}} dt\,
\mu_k(t)\, m_a$.  We will see that this results in a good
approximation to the actual evolution as determined by direct numerical
solution of the mode equation (\ref{axfluc}).

\subsection{\label{expansion}Resonant  Solution with Expansion}

\begin{figure}[t]
\centering
 \hskip -0.5 cm
\leavevmode\epsfysize= 5.5cm \epsfbox{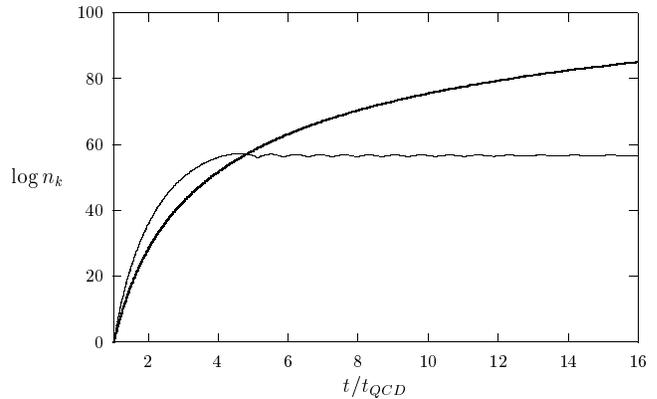}\\
\vskip  0.2cm
 \caption[fig2a]{\label{fig:fluc1}
The $\log$ of the
axion fluctuations $\vert \phi_k \vert^2$
in an expanding  
universe as a
function of time (in units of $t_{QCD}$).  We plot the maximally  
excited mode,
$k \approx {1 \over 5} \theta_o m_a$ for $\theta_o = 0.1$
(heavy line) in an expanding universe and the mode
$k = {1 \over \sqrt{8}} \theta_o m_a$ that would be maximally excited in a
static universe (light line).  As discussed in the text, the upper edge of
the instability band shifts beneath these modes after a time
$t_k = ( \frac{\theta_o m_a}{2 k} )^4 t_{QCD}$, which is $42 \times t_{QCD}$
and $4 \times t_{QCD}$ for the modes shown. }
\end{figure}

  We can now incorporate the expansion of the universe.
Fortunately, this also can be done analytically, because
the period of axion fluctuations $\sim m_a^{-1}$ is much smaller than the
Hubble time $H^{-1}$ around $t_{QCD}$.
We incorporate expansion into our static universe results by
including the redshifting of the wavelength, $k \to k/a$, and the
dilution of the background field, $\theta_0^2 \to \theta_0^2/a^3$.
Now, instead of expression (\ref{simple}), we have the time-dependent
characteristic exponent
\begin{equation}\label{5}
\mu_k(t)={k \over 4 m_a a}\sqrt{ {\theta_0^2 \over   a^3} -
  {4 k^2 \over m_a^2a^2}} \ .
\end{equation}
 From here it follows that  the expansion of the universe  can  shift
the resonant modes  out of the resonance. However, as the
restructuring of the band via expansion is somewhat faster than
the redshifting of modes, the modes actually
will be blueshifted relatively to the instability zone.
That is,  the upper
(larger momentum) edge of the instability band will shift toward lower
momentum values faster than the modes are redshifted down.
From  (\ref{5}) one can immediately estimate the time $t_k$ during  
which a given mode will be amplified,
$t_k = \left( {{\theta_0 m_a} \over {2k}} \right)^4 t_{QCD}$.
In Fig.~\ref{fig:fluc1} we plot the evolution of fluctuations as a function
of time, obtained by a numerical integration of Eq.~(\ref{axfluc}),
which confirms the analytical estimation of $t_k$.

\begin{figure}[t]
\centering
 \hskip -0.5 cm
\leavevmode\epsfysize= 5.5cm \epsfbox{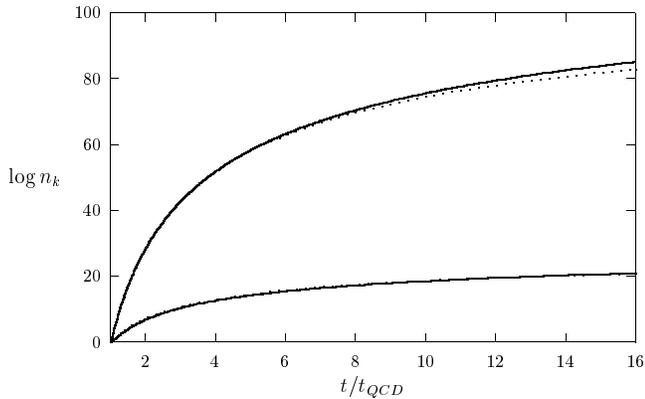}\\
\vskip  0.2cm
 \caption[fig2a]{\label{fig:fluc2}
The $\log$ of the axion fluctuations
$\vert \phi_k \vert^2$
 in an expanding  
universe as a
function of time (in units of $t_{QCD}$).  We plot the maximally  
excited mode,
$k \approx {1 \over 5} \theta_o m_a$, for $\theta_o = 0.1$
(upper curves) and $0.05$ (lower curves).  The solid lines are from an
exact numerical integration of the mode equation (\ref{axfluc}) in an
expanding universe.  The dotted curves are the approximation:
$\ln{\vert \phi_k \vert^2 }
 \approx {2 \int^{t}_{t_{QCD}} dt~ \mu_k(t) m_a}$; the
integral is given by formula (\ref{integral}) evaluated for an arbitrary
$t > t_{QCD}$.}
\end{figure}

 Amplification of fluctuations  by the resonance is determined by the
number of e-foldings:
 $\ln \vert \phi_k  \vert \simeq\int^{t_k}_{t_{QCD}} dt \mu_k(t) m_a$,
where we have assumed
no back reaction for the moment. From  Eq.~(\ref{5}) we find
\begin{eqnarray}\label{integral}
&&\int^{t_k}_{t_{QCD}} dt~ \mu_k(t)\,  m_a \nonumber\\
&&=  {\theta_0^2 \over {2 a_k}} \left(
\sqrt{a_k - 1}
- \arctan{\sqrt{a_k - 1}} \, \right)(m_a t_{QCD}),
\end{eqnarray}
where, $a_k$ is the change in scale factor from $t_{QCD}$ to $t_k$.
Specifically,
\begin{equation}\label{scale}
a_k = \sqrt{ t_k \over t_{QCD} } = \left({{\theta_0 m_a} \over {2 k}}
\right)^2.
\end{equation}
Expression (\ref{integral}) reaches its  maximal value \\
$ \int^{t_k}_{t_{QCD}} dt~ \mu_k(t) \approx 0.091~ (m_at_{QCD})  
\theta_0^2$
for $k \approx {1 \over 5} \theta_0 m_a$.
Thus, the leading amplification of the amplitude of a fluctuation
is given by
\begin{equation}\label{ampl}
\phi_k \simeq 10^{ 0.039  (m_a t_{QCD})  \theta_0^2} \ .
\end{equation}
This formula is the main technical result of the paper regarding
the axionic resonance.
The strength of the  resonance depends on two numerical
factors: the effective number of axion oscillations within a
Hubble time around the QCD epoch, $m_a t_{QCD}$, and the
value of the axion misalignment angle at $t_{QCD}$, $\theta_0$.
In Fig.~\ref{fig:fluc2} we plot the time evolution of the axion
fluctuations obtained with the numerical solution of Eq.~(\ref{axfluc})
together with that derived with analytic formula (\ref{integral})
for $\int _{t_{QCD}}^t dt \mu_k(t) m_a$ for arbitrary $t$.
This demonstrates the reliability of our analytic anzats.

\subsection{\label{res}Efficiency of the Axion  Resonance }

Let us estimate the efficiency of the axion resonance in Eq.~(\ref{ampl})
for a viable range of  parameters.
For the axion mass range $m_a \simeq 10^{-3}-10^{-4}$ eV and
$ t_{QCD} \simeq 7.7 \times 10^{-6}$ sec,
we have $m_a t_{QCD}  \approx
 5.85 \times 10^{6}  \left( { m_a \over {5 \cdot10^{-4} \hbox{\ eV} }}
\right)$.
Despite this large value for the number of axion oscillations in a Hubble
time, the amplification of axion fluctuations is not generally very
large because of the small average misalignment angle, $\theta_0$,
estimated by formula (\ref{omega1}).
However, as was mentioned at the beginning of this section, in most
interesting cases the background  field
has large scale inhomogeneities.  Therefore, we expect to have
regions where $\theta_0$ exceeds the estimation (\ref{omega1}).
In this case, only the r.m.s. value of $\theta_0$ is given by (\ref{omega1}).

	To amplify quantum axion fluctuations to the level
where fluctuations are comparable with the background field
$ \langle \delta \phi^2  \rangle \sim \phi_0^2$,  the exponent in  
(\ref{ampl})
should be as large as
$0.039 ~ (m_a t_{QCD}) ~ \theta_0^2 \gtrsim  23$. Thus,
regions of the universe where the background phase exceeds the level
\begin{equation}
\theta_0 \gtrsim 10^{-2}   \left({ m_a \over {5 \cdot 10^{-4}
 \hbox{\ eV} }}\right)^{-1/2}
\label{level}
\end{equation}
will have significantly amplified quantum fluctuations of axions.
Additionally, it has been noticed  \cite{ktk} that nonlinear axion  
dynamics
prior to $t_{QCD}$ will further enhance the amplitude in the initially
high $\theta_0$ regions.
If there are small scale, classical axion fluctuations,
the threshold value of $\theta_0$ for the effective resonance will be even
lower.  The  wavelength of the primary resonant mode,
$k \approx {1 \over 5} \theta_0 m_a$,  is significantly
smaller that the scale, $L$, over which $\theta_0$ is expected
to be inhomogeneous
\begin{equation}
  L k \sim 10^{-2} \theta_0~ t_{QCD} m_a \gg 1 \ .
\label{scale}
\end{equation}
The region of excessive $\theta_0$ corresponds to a proto-minicluster.
The larger $\theta_0$ is, the stronger the resonance effect will be.
This means that in regions of particularly large $\theta_0$,
the  background field will decay
into axion fluctuations within several Hubble times $\sim O(10)t_{QCD}$.
The resulting axion fluctuations  can be viewed
as a superposition of very slowly moving (non-relativistic) axion
waves of wavelength $k^{-1}$. Thus, parametric resonance will cause
the (quasi)homogeneous field in the region of
a proto-minicluster to become strongly inhomogeneous.
Then, when axions come to gravitationally dominant,
a large number ($\sim (L k)^3$) of dense axion clumps
will form from the remnants of the proto-minicluster.

We have assumed that the background
fluctuations in the region of
proto-minicluster  are harmonic and coherent. The actual  
fluctuations may not
be
perfectly coherent.  In this case,
one would expect the resonance to be significant if $\theta_0$
exceeds a level which is somewhat greater than the idealized estimation
(\ref{level}). To study the resonance in the case when  the background
fluctuations are not perfectly coherent, one could use
lattice simulations like those in \cite{ktk} but with much
greater resolution.  Alternatively, one could try to mimic the background
fluctuations as a random one dimensional process $\phi(\tau)$.  These are,
however, beyond the scope of the current paper.
Also, we cannot extend our results to the scenarious
where axions at $t_{QCD}$ are relativistic \cite{walls}. 
For now, we consider our results as an indication of possible
strong axion resonance in the regions of proto-miniclusters.

\section{\label{stability}Preheating in Natural Inflation    }

Another important application of the theory Sine-Gordon parametric
resonance elaborated above is to particle production
in the Natural Inflation scenario \cite{natural},
 where the pseudo-Nambu-Goldstone
boson $\phi(t)$ plays the role of the inflaton field with the potential
(\ref{1}). The typical parameters of Natural Inflation
are $f \simeq M_p$, $m_{\phi} \simeq 10^{12}$ GeV, and $\Lambda  
\simeq 10^{15}$
GeV.
First, we consider the rate of creation of inflaton quanta
due to the self-interaction. This problem is reduced to the
Sine-Gordon parametric resonance in an expanding universe.
Then we consider other channels of inflaton decay.
Although that issue goes beyond the stability of fluctuations
in the Sine-Gordon  model, we add this to the paper to complete
the picture of reheating and preheating in the Natural Inflation
scenario.

\subsection{\label{stability2}Self-interacting Inflaton  Decay}

The homogeneous inflaton field   $\phi(t)$    in this model
obeys the equation
\begin{equation}\label{SGI}
\ddot \phi + 3H\dot \phi+   m_{\phi}^2 f \sin {\phi \over f}=0 \ .
\end{equation}
During the inflationary stage, the field  $\phi(t)$
is slowly rolling down to the minimum of the potential
(\ref{1}). After inflation, the field oscillates about the
minimum of the potential with a frequency $m_{\phi}$.
After several cycles, these oscillations  can be  simply
approximated by harmonic ones, $\phi(t) \approx \Phi(t) \sin m_{\phi}t$,
with decreasing amplitude,
$\Phi(t)   ={ \Phi_0 \over {a^{3/2}}} $.
The  initial amplitude $\Phi_0 \simeq  O(10^{-2})M_p$.
 Inflaton oscillations correspond to the matter dominated equation  
of state for which $a(t) \propto t^{2/3}$.

Let us now turn to the vacuum inflaton fluctuations, $\delta \phi$.
The temporal part of an eigenmode, $\phi_k(t)$, obeys the equation
\begin{equation}\label{inffluc}
\ddot \phi_k + 3H\dot \phi_k + \left({k^2 \over a^2}
 +m_{\phi}^2 \cos \theta \right) \phi_k=0 \ ,
\end{equation}
where the oscillating background phase $\theta={{\phi(t)} \over f}$
(c.f. Eq.~(\ref{axfluc})).
At the start of oscillations,
the typical expansion time, $H^{-1}$,
is comparable with the period of oscillations and
exceeds it at the latest stages.  We find that
the parametric resonance of inflaton fluctuations
is insignificant.  This result can be understood as follows.
Resonance occurs in a single resonance band which spans momenta
$k/a(t)$ between
zero and $\sim \theta_0(t) m_{\phi}$, where 
$\theta_0(t) = {\Phi_0 \over f} = \Theta_0 a^{-3/2}$.
For this reason, resonant modes
will not be redshifted out of the band,
as was suggested in \cite{df}. Just the opposite occurs;
momenta redshift more slowly than the background field amplitude, and
they become blueshifted relative to the resonance band, just as occurred in
the axion case, see Sec.~\ref{expansion}.
Let us assume for a moment that the expansion is slow.
Then we can use the approximation (\ref{5}) for the characteristic
exponent.  The the number of e-foldings $\int dt~\mu_k(t) m_{\phi}$
will again be proportional to $m_\phi t_0 \theta_0^2$, where $t_0$ is the
start of oscillations.  However, the factor 
$m_\phi t_0$ is of order one, not $10^6$ as it was in the axion case.
Also, the phase $\theta$ is necessarily rather small.
The net result is, at best, a few e-foldings.
Now, as the actual  expansion at the beginning of the oscillatory phase
is not slow, the resonance will be even less  efficient.
Thus, we confirm the claim of \cite{df} that the parametric resonance
of the inflaton field due to the self-interaction in Natural Inflation
is insignificant, although our explanation is somewhat different.

\subsection{\label{stability2}  Resonant  Inflation Decay
 with a $g^2 \phi^2 \chi^2$ Interaction}

However, to make the picture of reheating in the
Natural Inflation scenario more comprehensive, we have to consider
other channels of inflaton decay.
Let us consider the decay of the inflaton into
other bosons $\chi$, via a coupling ${1 \over 2}g^2 \phi^2 \chi^2$.
Using a decomposition of the  Bose field $\chi$ into
eigenfunctions $\chi_{k}(t)\, e^{ -i{{\bf k}} \cdot {{\bf x}}}$
with comoving momentum ${\bf k}$, we obtain, for their temporal
parts,
\begin{equation}\label{bosons}
\ddot \chi_k + 3H\dot \chi_k + \left({k^2 \over a^2}
 +g^2 \phi(t)^2 \right) \phi_k=0 \ .
\end{equation}
where   the background oscillations $\phi(t)$  are given by the  
solution of
Eq.~(\ref{SGI}).

  Eq.~(\ref{bosons}) is an oscillator-like equation
with a periodic effective frequency. Therefore, we should expect
resonant solutions $\chi_k \sim \exp \int dt \mu_k(t) m_\phi$.
Indeed, when the amplitude of the background phase is sufficiently low,
the background oscillations correspond to those in a massive
theory ${1 \over 2} m_\phi^2 \phi^2$. Therefore, Eq.~(\ref{bosons})
is similar to the well-studying case of preheating in
 ${1 \over 2} m^2 \phi^2$-inflation \cite{KLS97}.
We introduce the usual resonance parameter $q={{g^2 \Phi^2} \over m^2}$.
If $q \gg 1$, we will have the creation of $\chi$-particles in the
 broad resonance regime. However, due to the expansion of the universe
there will be no distinct resonance bands, but rather a broad range
of resonant momenta $\Delta k \simeq q^{1/4}m_\phi$ where the 
resonance occurs in a stochastic manner \cite{KLS97}.
The growth of the
particle's occupation number $n_k(t)$ will be a stochastic process.
For moderate values of $q$ we approach the regular
resonance regime.

\begin{figure}[t]
\centering
 \hskip -0.5 cm
\leavevmode\epsfysize= 5.5cm \epsfbox{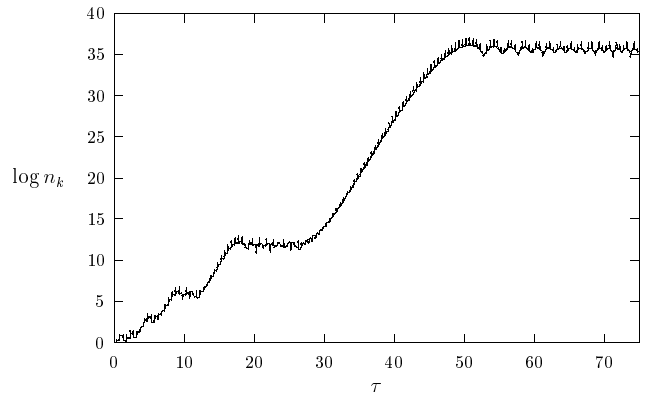}\\
\vskip  0.2cm
 \caption[fig2a]{\label{fig:fluc3}
The typical  stochastic resonant production of particles at the
particular choice of comoving momentum $k^2_0 = 0.1 m_{\phi}^2$,
and $q = 10^2$ with expansion.  Time
is measured in the number of inflaton oscillations since the end of
inflation.}
\end{figure}

	Let us estimate the range of allowed $q$ values.
From the basic parameters of the Natural Inflation scenario, 
we have $q={{g^2 \Phi_0^2} \over m_\phi^2} \simeq g^2 10^{10}$.
The coupling parameter $g^2$ is restricted by the condition that radiative
 corrections do not alter the potential (\ref{1}),
${ 1 \over 64\pi^2} g^4 \phi^4 \log {\phi \over m}  \leq \Lambda^4  $.
This gives us $g \leq { 5\Lambda  \over f} \sim 5 \cdot 10^{-4}$,
and finally $q \leq 10^3$.
In  Fig. \ref{fig:fluc3} we show the evolution of the number density
$n_k(t)$ of $\chi$-particles calculated from a numerical solution to
the equation for fluctuations (\ref{bosons}).
Thus, for the  resonance parameter $1 \leq q \leq 10^{3}$
the parametric resonance is very effective,
 and  leads to the preheating of bosons $\chi$.
The leading channel for inflaton energy transfer
will be that with the greatest $q$.

  Additionally, it was found very recently that
there can be preheating of fermions $\psi$ due to an
inflaton coupling $h\bar  \psi \phi  \psi$ \cite{GK}.
The creation of fermions occurs in the regime of parametric excitation,
which is very different from the perturbative calculations used
earlier \cite{df} for Natural Inflation.
In  perturbative reheating, the reheat temperature was found to be 
rather low, $T_r \simeq g 10^8$ GeV \cite{natural}, where
$g^2$ is coupling between inflatons and other particles.
However, as we have seen, the preheating of bosons and fermions
in Natural Inflation can be very efficient.
Therefore, there it is a possible to have a
larger value of  the reheating temperature in this 
scenario than that calculated with the perturbation theory.

\section{\label{disc} Summary }

In this paper, we have discussed the general theory
of parametric resonance of fluctuations around an
oscillating homogeneous scalar field with a
Sine-Gordon potential. We found a closed form analytic
resonance solution which shows that, in such a theory, there is a single 
instability band which starts at zero momentum. 
The characteristic exponent and width of the resonance band 
are strongly dependent on the dimensionless amplitude, or
misalignment angle, $\theta_0$, of the background  
oscillations, decreasing as  $\theta_0$ decreases.
Although the background oscillations are well described 
by a massive theory in the regime of small $\theta_0$,
the correct properties of the parametric resonance in this regime
cannot be obtained by making the small angle approximation in 
the equation for fluctuations, which would lead to a Mathieu equation.
This is due to the non-perturbative nature of the
parametric resonance.

As a particular application of the general theory of
Sine-Gordon parametric resonance, we considered the
stability of axion fluctuations in the cosmological
model with cosmic axions. By the QCD epoch, axions
are non-relativistic and can be considered as a
coherently oscillating background field with a misalignment angle 
$\theta_0$ and frequency $m_a \sim 10^{-4}$ eV.
The expansion of the universe is important to the strength of the
resonance, but is not itself deadly as the typical expansion time 
significantly exceeds the period of axion fluctuations, e.g. 
$m_a t_{QCD} \gg 1$.
If the amplitude of fluctuations is as small as $\theta_0 \leq 10^{-3}$,
as required of the r.m.s. value in an axion dominated cosmology,
then the expansion of the universe overdominates for viable ranges of $m_a$
and axionic parametric resonance is inefficient.
However, there are axion models for which the background field  
$\theta_0({\vec x})$ at the QCD epoch is not perfectly
homogeneous. If there were no resonance, the regions of large $\theta_0$
would become axion miniclusters when axions come to gravitationally dominant.
Assuming the background oscillations are perfectly coherent
in the regions of proto-miniclusters,
we found that the axion fluctuations are unstable in the
proto-minicluster regions where $\theta_0 \gtrsim 10^{-2}$
for $m_a = 5 \times 10^{-4}$ eV.
The wavelengths of the resonant modes are significantly smaller
than the expected scale of proto-minicluster regions.
Thus, in these regions background axions at rest will resonantly  
decay into moving  non-relativistic axions. In other words, a smoothly
varying axion field $\theta_0({\vec x})$ will decay into
a strongly inhomogeneous axion field.
As a result, at the epoch when axions become gravitationally dominant,
instead of a single, massive axion minicluster we will have a minicluster
composed of a large number of smaller and denser axion 
``mini-miniclusters''.
This may be important for axion detection strategies.
The effect of axion resonance in the regions of miniclusters
should be checked with lattice simulations or with a more 
sophisticated analytic treatment of the background oscillations
which may not be perfectly coherent.
Therefore, we consider our calculations as an indication
that there may be strong resonance effects in axion miniclusters.

Another application of the general theory of the
Sine-Gordon parametric resonance is to particle production in
the Natural Inflation scenario, where the inflaton field
has a Sine-Gordon self-interaction.
We considered the production of inflaton fluctuations
and found that the  resonant decay of inflatons is not effective.
Beyond the  Sine-Gordon self-interaction, we consider the
inflaton decay in this scenario via a four-leg interaction with
other bosons.  The channels of decay for which the coupling parameter
is in the region $10^{-5} \leq g \leq 5 \cdot 10^{-4}$ will
lead to a broad parametric resonance; smaller $g$
will correspond to a narrow  resonance.
 Together with the preheating of fermions, 
this opens the possibility of preheating in the Natural Inflation scenario.

\bigskip
\section*{Acknowledgments}
The authors are grateful to A.~Linde for valuable discussions
and suggestions and to I. Tkachev for useful comments.
  L.K. was supported in part by NSF grant AST95-29-225.
A.S. was supported  in part by the Russian Foundation
for Basic Research, grant 96-02-17591, and the Russian
Research Project ``Cosmomicrophysics''.

\section*{Appendix  A: Background Solution of the Sine-Gordon  Equation}

Here we derive the general  homogeneous solution  of the
Sine-Gordon equation (\ref{SG})
for an arbitrary initial amplitude $\phi_0$,  not necessarily much
smaller than $f$,
with no expansion of the Universe taken into account.
Without loss of generality, we can restrict ourselves to the finite
 motion $\vert \phi(t) \vert \leq
\pi f$, with the constant  total energy $E$ given by
(\ref{energy}).
Let us put $y(t) = \tan {\phi(t) \over 2f}$. Then from  the
 energy integral (\ref{energy})
one finds
\begin{equation}\label{energy1}
\dot y^2= { 1 \over  2f^2} (1+ y^2) \left( E-(2m_a^2f^2-E)y^2\right) \ .
\end{equation}
Let us now recall the equation for the elliptic
cosine, $C\equiv cn (\tau; {\cal K})$,
 as a function of some variable $\tau$ where $ {\cal K}$ is its modulus:
\begin{equation}\label{ell}
\dot C^2=  {\cal K}'^2+ ({\cal K}^2-
 {\cal K}'^2)C^2-{\cal K}^2C^4 \ , \,   {\cal K}'^2=1- {\cal K}^2 \ .
\end{equation}
Here, $  \dot{()} $ stands for the $\tau-$derivative.
It is easy to check that Eq.~(\ref{energy1})
can be reduced to an equation for the elliptic
 cosine by means of the substitutions $y= \sqrt{ { E \over  
{2m_a^2f^2  -E}}}
cn(\tau; {\cal K})$, $\tau=m_at$ where a dimensionless parameter is
$\epsilon=E/m_a^2f^2$ and the modulus is
 ${\cal K}= \sqrt{\epsilon/2}$.
Therefore,  for the  background oscillations we have
$\tan {\phi(t) \over 2f}= \sqrt{\epsilon \over {2-\epsilon}}
cn( m_a t; \,   {\cal K})$, which finally results in
Eq.~(\ref{hom}).
The (dimensionless) period of the background fluctuations,
$T_{\tau}$, is equal to
\begin{eqnarray}
&&T_{\tau}=\sqrt{8}\int\limits_0^{\arccos (1- \epsilon )}
{d {(\phi \over f)} \over \sqrt{\epsilon-1+\cos{(\phi \over f)}}}
 \nonumber\\
&&=\sqrt{8} \int\limits_0^{\pi/2} {d \vartheta \over
\sqrt{ (1+\sin^2 \vartheta)\Bigl(1+
(1- \epsilon)\sin^2 \vartheta \Bigr)}} \ ,
\label{period}
\end{eqnarray}
which is equal to $4{\bf K}\Bigl( \sqrt{ \epsilon \over 2}\Bigr)$.

\section*{Appendix  B:  Algebraic form of the Lam\'{e}  Equation}

Here we demonstrate, for the sake of completeness, that the equation for
fluctuations $\phi_k(t)$ in the Sine-Gordon model  (\ref{fluc3})
(without expansion) can be reduced to the Lam\'{e} equation
with  a particular choice of  the coefficients. Indeed, the
general algebraic form of the  Lam\'{e} equation
 is \cite{ince}
\begin{eqnarray}
&&Y''+ {1 \over 2} \left({1 \over z-e_1}+
{1 \over z-e_2}+{1 \over z-e_3}  \right) Y' \nonumber\\
&&+ { {Az+B} \over 4(z-e_1) (z-e_2) (z-e_3)} Y=0 \ ,
\label{lame}
\end{eqnarray}
where $Y(z)$ is the solution of the  Lam\'{e} equation.
The equation for fluctuations  (\ref{fluc3}) has the form  (\ref{lame})
with $e_1=1$, $e_2=1-\epsilon$, $e_3=-1$, $A=-2$, $n=1$,
$B=\lambda=-2\kappa^2$.
Therefore,  $\phi_k(t)$ is a solution of the  Lam\'{e} equation
with these parameters.
Notice that the Lam\'{e} equation has already been met in the context of
parametric resonance  in the theory of preheating with a $\lambda \phi^4$
effective potential \cite{GKLS}.

\section*{Appendix  C:  Derivation of the Characteristic Exponent }

Here we   show how one can derive Eq.~(\ref{mu2})  for the
characteristic exponent $\mu_k$ from the analytic solution (\ref{form1}).
Eq.~(\ref{form1}) describes both solutions, $\phi_1(z)$ and $\phi_2(z)$.
 The resonant
solution $\phi(z)$ consists of four monotonic parts within
a single period of the background oscillation.
 It turns
out that at different quarters of the period either  $\phi_1(z)$
 or  $\phi_2(z)$
corresponds to the exponentially growing solution.   Indeed, the  
square of the
resonant solution within the first quarter of a period, while
 $z$ is decreasing from $1$ to $(1-\epsilon)$, is
\begin{equation}\label{A1}
\phi^2(z) = \phi^2_0\exp\left[\int\limits_1^z{dz\over M(z)}
\Bigl(1+{c_k\over\sqrt{(1-z^2)(z-1+\epsilon)}}\Bigr) \right] ,
\end{equation}
where $M(z)$ is  given by (\ref{deg}),   $ c_k $ is given by  
(\ref{C}), and
$\phi_0^2$ is the square of the resonant solution in the beginning of
 the period
when $z = 1$.

Within the second quarter  of the period,
while
 $z$ is increasing from $(1-\epsilon)$  to $1$,
one has
\begin{equation}\label{A2}
\phi^2(z) = \phi_{1/4}^2\, \exp\left[ \int\limits_{1-\epsilon }^z{dz\over
M(z)}\Bigl(1 -  {c_k\over \sqrt{(1-z^2) (z-1+\epsilon)  }}   
\Bigr)\right] \ ,
\end{equation}
where $\phi_{1/4}$ is the value of  $\phi_{z}$ after the first
quarter of the period
$\phi_{1/4} \equiv \phi(z=1)$.

Then the value  of $\phi^2$ after   half of a period is
\begin{equation}\label{A3}
 \phi_{1/2}^2
 =  \phi_0^2
 \exp\left[-2c_k \int\limits_{1-\epsilon }^1{dz \over
{M(z) \sqrt{(1-z^2)(z-1+ \epsilon)  }  }  } \right] \ ,
\end{equation}
where the integral is understood as its   principal value.
The resonant solution has the generic form $\phi(z(\tau)) =
 P(z(\tau)) e^{\mu \tau}$, where
$P(z)$ is a periodic function. Since $P$ has a period equal to    
half of the
period of the inflaton oscillation, Eq.~(\ref{A3}) is sufficient to  
find $\mu$:
\begin{equation}\label{A4}
{\mu T_{\tau}\over 2} = - c_k \int\limits_{1-\epsilon }^1{dz \over
{M(z) \sqrt{(1-z^2)(z-1+ \epsilon)  }  }  }      > 0 .
\end{equation}
The integral in this equation can be reduced to $I(\kappa^2)$ given by
(\ref{I}).
\begin{eqnarray}\label{A5}
&&  - \int\limits_{1-\epsilon }^1{dz \over
{M(z) \sqrt{(1-z^2)(z-1+ \epsilon)  }  }  }=\nonumber\\
&&= 2\int\limits_0^{\pi/2} {{d \vartheta}
 \over {\sqrt{ (1+\sin^2 \vartheta)\Bigl(1+
(1-  \epsilon)\sin^2 \vartheta \Bigr)} }} \times \nonumber\\
&&{{\sin^2 \vartheta} \over
{ 1+\Bigl(1-\epsilon +2\kappa^2\Bigr)\sin^2 \vartheta} }
\equiv I(\kappa^2).
\end{eqnarray}

\end{document}